\documentclass{aa}  

\usepackage{graphicx}
\usepackage{xcolor}
\usepackage{booktabs}
\usepackage{txfonts}
\usepackage{url}
\usepackage[normalem]{ulem}

\usepackage[pdfencoding=auto,psdextra]{hyperref}
\hypersetup{
    colorlinks=true,
    linkcolor=blue,
    filecolor=magenta,      
    urlcolor=blue,
    citecolor=blue
}
\defcitealias{Sartoris2020}{S20}
\defcitealias{Biviano23}{B23}

\begin{document}

   \title{Testing Refracted Gravity with kinematics of galaxy clusters}

   \author{L. Pizzuti\inst{1} \and F. Fantoccoli\inst{1} \and V. Broccolato\inst{2} \and A. Biviano\inst{3,4} \and A. Diaferio\inst{2,5}}


\institute{Dipartimento di Fisica G. Occhialini, Universit\`a degli Studi di Milano-Bicocca, Piazza della Scienza 3, I-20126 Milano, Italy\\
    \email{lorenzo.pizzuti@unimib.it}
    \and
    Dipartimento di Fisica, Universit\`a di Torino, via P. Giuria 1, 10125, Torino, Italy
    \and INAF-Osservatorio Astronomico di Trieste, via G. B. Tiepolo 11, I-34131, Trieste, Italy
    \and
    IFPU-Institute for Fundamental Physics of the Universe,
    via Beirut 2,
    34014 Trieste, Italy
    \and
    Istituto Nazionale di Fisica Nucleare, Sezione di Torino, via P. Giuria 1, 10125, Torino, Italy
    }



   \date{Received October ??, 2024; accepted ??}


  \abstract{Refracted Gravity (RG) is a  a classical theory of gravity where a gravitational permittivity $\epsilon(\rho)$, a monotonically-increasing function of the local density $\rho$, is introduced in the Poisson equation to mimic the effect of dark matter at astrophysical scales. We use high precision spectroscopic data of two massive galaxy clusters, MACS J1206.2-0847 at redshift z=0.44, and  Abell S1063 (RXC J2248.7-4431) at z=0.35, to determine the total gravitational potential in the context of RG and to constrain the  three, supposedly universal, free parameters of this model. Using an upgraded version of the \textsc{MG-MAMPOSSt} algorithm, we perform a kinematic analysis which combines the velocity distribution of the cluster galaxies and the velocity dispersion profile of the stars within the Brightest Cluster Galaxy (BCG). The unprecedented dataset used has been obtained by an extensive spectroscopic campaign carried out with the VIMOS and MUSE spectrographs at the ESO VLT. We found that RG describes the kinematics of these two clusters as well as Newtonian gravity,  
  although the latter is slightly preferred. However, (i) each cluster requires a different set of the three free RG parameters, and (ii) the two sets are   
  inconsistent with other results in the literature at different scales. 
  We discuss the limitation of the method used to constrain the RG parameters as well as possible systematic effects which can give rise to the observed tension, notably deviations from the spherical symmetry and from the dynamical equilibrium of the clusters.  
  }


    \authorrunning{Pizzuti et al.}
    \titlerunning{Testing Refracted Gravity with clusters}
    
   \maketitle
%
\section{Introduction} \label{sec:intro}
The standard cosmological model ($\Lambda$CDM) is based on General Relativity (GR) to describe the gravitational interaction. In this framework, a cosmological constant $\Lambda$ is introduced in  Einstein's field equation to explain the accelerated expansion of the universe (\citealt{Reiss01,perlmutter99}); moreover a non-baryonic Cold Dark Matter (CDM) component \citep{Dodelson96}, which should account for about  30\% of the cosmological mass budget, is required to explain the formation and evolution of cosmic structures at different scales (e.g. \citealt{SartorisDM,Salucci2019,Arbey2021,Limousin22}). 

The predictions of the $\Lambda$CDM paradigm  agree with a large variety of independent observations made in recent years (\citealt{Planck2020}); nevertheless, a quite large number of unsolved critical issues challenge $\Lambda$CDM both at cosmological and astrophysical scales (see e.g. \citealt{Perivolaropoulos_2022} for a review).

In the last decades, several suggestions have been proposed to alleviate the tensions between the predictions of the standard model and the observations. Among the various alternatives and extension of $\Lambda$CDM, one popular possibility is to reshape the theory of GR in such a way that the observed effects of the dark sector are mimicked by a modified gravitational interaction (e.g. \citealt{Famaey_2012, Joyce16, Shankaranarayanan_2022} and references therein). The large variety of modified gravity (MG) models has been extensively tested by means of several probes from the Solar System to cosmological scales (e.g. \citealt{Tereno_2011,Braglia_2021,Boumechta:2023qhd,Banik_2023}).  In this paper we focus on the model of \textit{Refractive Gravity} (RG), introduced by \cite{matsakos2016dynamics}, a phenomenological extension of standard gravity without dark matter (DM), where the Poisson equation is modified by introducing a \textit{gravitational permittivity} $\epsilon(\rho)$, a monotonic function of the mass density $\rho$. The role of the gravitational permittivity is to enhance the gravitational field in regions where the density goes below a critical value,  producing the same effect of DM at the scales of galaxies and galaxy clusters (see also \citealt{Cesare2024} for a review), in analogy with the modification of electric field lines in contiguous different media in classical electrodynamics.
For example, for flat systems (as disk galaxies,  \citealt{matsakos2016dynamics, RG2020}), the presence of $\epsilon(\rho)$ in the Poisson equation produces a bending in the gravitational field lines of the system causing an increment of the gravitational pull through the plane parallel to the galactic disk.
In spherical systems, the field lines remain radial, but the intensity of the gravitational field is boosted by the inverse of the permittivity function.

Whereas the RG framework has been derived from empirical considerations, there has been a recent attempt to extend the model towards a covariant formalism  as a scalar-tensor theory of a gravity \citep{RGcovariant}. In this covariant formulation, a scalar field plays the role of the gravitational permittivity; moreover, at large scale the same scalar field is also responsible for the accelerated expansion of the universe, providing a unified description of the whole dark sector. 

The RG model has been already tested on a sample of disk and elliptical galaxies (\citealt{RG2020,RG2023}), and some preliminary studies have been performed at a scale of galaxy clusters by \cite{matsakos2016dynamics}. In this context, clusters are an excellent ground to test GR and its modifications at the edge between astrophysical and cosmological scales. While many modified gravity models alternative to dark matter (e.g. MOdified Newtonian Dynamics, MOND, \citealt{Milgrom1983}) fail in reconciling lensing and kinematics determination of the total gravitational potential (see e.g. \citealt{Hodson17}), RG may offer a way out with a suitable choice of the model's free parameters.
In this work we aim at constraining those parameters by performing a kinematic analysis of two massive galaxy clusters, extensively studied within the Cluster Lensing And Supernova survey
with Hubble (CLASH; \citealt{Postman12}) and the spectroscopic
follow-up with the Very Large Telescope (CLASH-VLT; \citealt{Rosati14}) programs: MACS J1206.2-0847 (hereafter MACS 1206) at
redshift $z = 0.44$ and Abell S1063 at $z = 0.35$. By means of the \textsc{MG-MAMPOSSt} code of \cite{Pizzuti2021}, we use high quality imaging and spectroscopic data of projected positions and  velocity distribution of member galaxies, combined with the stellar velocity dispersion data of the brightest cluster galaxy (BCG),  to reconstruct the total mass profile in RG by accurately modelling all the baryonic components.

The article is organized as follows: in Sec. \ref{sec:theo} we briefly review the basic equations of RG and we derive the expression of the total gravitational potential used in our analysis. In Sec. \ref{sec:data} we describe the dataset of the two galaxy clusters used for the kinematic mass reconstruction. In Sec. \ref{sec:MAM} we provide a short overview of the \textsc{MG-MAMPOSSt} code and the modification we implemented for this study, further illustrating the set-up for the analysis.  Sec.~\ref{sec:results} is devoted to the presentation of the main results which are discussed in Sec.~\ref{sec:disc}, and compared with previous analyses at galactic scales. Finally, we draw our conclusions in Sec. \ref{sec:conc}. Throughout this paper, we adopt a flat cosmological background with $\Omega_m = 0.3$, $\Omega_{\Lambda} = 0.7$ and $H_0 = 70\,  \text{km}\,\text{s} ^{-1} \text{Mpc}^{-1}$.

\section{Theoretical Framework}
\label{sec:theo}
The \textit{gravitational permittivity} in classical RG modifies the gravitational field when passing from high to low density environments. This modification is analogous to what happens in classical electrodynamics where the electromagnetic field lines change crossing different contiguous media; a similar analogy can be made in optics, where a light ray is \textit{refracted} (from here the name of the theory). The modified Poisson equation in RG reads:
\begin{equation}
    \nabla \cdot [\epsilon(\rho) \nabla \Phi] = 4 \pi G \rho\, , \label{Poisson_1}
\end{equation}
where $\phi$ is the gravitational potential, $G$ the gravitational constant and $\epsilon(\rho)$ the permittivity function, considered as a monotonically-increasing function of the mass density $\rho$. It is important to notice that the permittivity does not need to be a function of the mass density alone, but it could in general depend on any other local variable that characterizes the source of the gravitational field (for instance, the energy density); however, consistently with previous work  (e.g. \citealt{RG2020,RG2023}), here we choose the simplest dependence on $\rho$.

The gravitational permittivity is defined as a function of mass density using a smooth transition function between two regimes, as follows:
\begin{equation} \label{eq:epsrho}
    \epsilon(\rho) = \epsilon_0 + (1 - \epsilon_0 ) \frac{1}{2} \left\{\tanh \left [\ln \left(\frac{\rho}{\rho_c}\right)^Q \right] + 1\right\}\, ,
\end{equation}
where $\epsilon_0$ is the gravitational permittivity in vacuum, $\rho$ represents the  local mass density  
and $\rho_c$ is a critical density above which standard gravity is recovered. In particular, in high density regions $\epsilon(\rho > \rho_c) = 1$, whereas in the low density regime $\epsilon(\rho < \rho_c) \xrightarrow{} \epsilon_0$. The steepness of the transition is controlled by the slope parameter $Q$. The phenomenological model of eq.~\eqref{eq:epsrho} allows all possible values of $\epsilon(\rho) \in [\epsilon_0, 1]$. 

The consequences of the presence of $\epsilon(\rho)$ as a modification of the Poisson equation are rather important when looking at the expansion of  equation \eqref{Poisson_1}:
\begin{equation}
    \frac{\partial \epsilon}{\partial \rho} \nabla \rho \cdot \nabla \phi + \epsilon(\rho) \nabla^2 \phi = 4 \pi G \rho \, . \label{Exp_poisson}
\end{equation}
For example, in flat systems (as disk galaxies) the first term on the left-hand side of equation \eqref{Exp_poisson} is non zero, and the gravitational field lines are \textit{refracted} parallel to the disk plane; therefore in the outermost regions of the disk the acceleration goes as $ r^{-1}$. This offers a whole new prospective in the analysis of rotational curves of disk galaxies, as pointed out by \cite{matsakos2016dynamics}.

In our study we are concerned with spherically symmetric systems; here the presence of the gravitational permittivity does not modify the direction of the field lines, but it increases the strength of the field by a factor $1/\epsilon(\rho)$ (e.g. \citealt{matsakos2016dynamics}). The gradient of the gravitational potential for a spherically symmetric galaxy cluster in classical RG is given by:
\begin{equation}
\label{eq:RG}
    \begin{split}
     &\frac{\text{d} \Phi}{\text{d}r} = \frac{G M_{\text{Bar}}(r)}{r^2} \times \\   
     & \left\{\epsilon_0 + (1 - \epsilon_0)\frac{1}{2}\left[\tanh \left(Q\,\ln \left( \frac{\rho_{\text{Bar}}(r)}{\rho_c}\right) \right) + 1 \right]\right\}^{-1}\,,\\
     \end{split}
\end{equation}
where $\rho_{\text{Bar}}(r) = \rho_{\text{gas}}(r)+ \rho_{\text{BCG}}(r)+ \rho_{\text{gal}}(r) $
is the total baryonic component, obtained as the sum of the gas contribution, the BCG  mass density and the density in stellar mass of the cluster galaxies. The corresponding total mass is obtained as 
\begin{equation}
\begin{split}
M_{\text{Bar}}(r) &= 4\pi \int_0^r{\text{d}x\,  \rho_{\text{Bar}}(x)x^2 } \\
& = M_{\text{gas}}(r)+ M_{\text{BCG}}(r)+ M_{\text{gal}}(r)\,.
\end{split}
\end{equation}
Note that one can re-write the right-hand-side of eq. \eqref{eq:RG} to mimic the standard Poisson equation by defining an \emph{effective} dynamical mass $M_\text{dyn} = M_\text{Bar}(r)/\epsilon(\rho)$.

\section{Dataset}
\label{sec:data}
We consider high-quality spectroscopic data of the galaxy clusters MACS 1206 at $z=0.44$ and Abell S1063 at $z=0.35$, which belong to a set of clusters extensively analysed within the CLASH \citep{Postman12} collaboration and its spectroscopic follow-up campaign CLASH-VLT \citep{Rosati14}, conducted with the VIMOS spectrograph. Additional spectroscopic datasets have been obtained using the MUSE integral field instrument in the cluster core \citep{Caminha2017}\footnote{The public redshift catalog and references
for these clusters are available at the \href{https://sites.google.com/site/vltclashpublic/home}{CLASH-VLT website}}

MACS 1206 appears to be an overall relaxed system, with only minor substructures in the projected distribution of the galaxies \citep{girardi15,Lemze01}. The dynamical relaxation condition is further suggested by the absence of strong deviation from Gaussianity of the line-of-sight (l.o.s. hereafter) velocity distribution (\citealt{Pizzuti:2022Vain}) and  nearly concentric structure of the projected mass components (see e.g. \citealt{UmetsuMACS}) .

Here we adopt the same sample of member galaxies as in \citet{Biviano23}, hereafter B23, which carried out a parametric multi-component reconstruction of the total mass profile from internal kinematic analyses. In particular, among the total of 3110 sources with measured reliable redshift (see \citealt{Balestra_2016} for a detailed description of the spectroscopic catalogue) in the cluster field, we identify 468  cluster members up to $R \sim 2.16\, \text{Mpc}$ using the \textsc{CLUMPS} selection algorithm (Appendix A of \citealt{Biviano21}), assuming as the cluster center the position of the BCG.

In addition, we consider six points for the BCG stellar velocity dispersion profile (VDP), in the radial range $[1,50]\,\text{kpc}$ from the center, obtained with the method of \cite{Sartoris2020}. The surface brightness profile is modeled as a Jaffe profile,
\begin{equation}
\label{eq:jaffe}
    L_J(r) = L_{\text{BCG}}\,\frac{r}{r_J}\left(1+ \frac{r}{r_J}\right)^{-1}\,,
\end{equation}
with $L_{\text{BCG}} = 4.92 \times 10^{11}\,\text{L}_\odot$, as measured in the $I$-band, and 
$r_J = 39 \,\text{kpc}$ (see \citetalias{Biviano23} and references therein).
The mass of the Intra-cluster Medium (ICM) hot gas is also taken from \citetalias{Biviano23}, and it is obtained from the analysis of Chandra ACIS-I exposures, using the method described in \cite{Ettori10}; the stellar mass of the satellite galaxies $M_*(r)$ is derived from the analysis of \cite{Annunziatella2014}.

For Abell~S1063 \citep{Abell89}, the combined datasets from the VIMOS and MUSE spectrographs consist in 3850 observed redshifts, from which a sample of 1234 cluster members was selected. 
The total spectroscopic sample of this cluster at redshift $z=0.346$ is presented in  \citet{Mercurio_2021} (see also \citealt{Caminha16}).
With this unprecedented dataset, \cite{Sartoris2020} - hereafter S20 - reconstructed the multi-component kinematic mass profile of Abell~S1063 from $\sim 1$~kpc up to the virial radius. As for the diffuse gas mass profile, $M_{gas}(r)$ and the mass in galaxies $M_\text{gal}(r)$ other than the BCG, we use the same profiles as in \citetalias{Sartoris2020}. In particular, they obtained the gas mass from the X-ray luminosity  archival Chandra ACIS-I observations, while $M_\text{gal}$ comes from the Spectral Energy Distribution (SED) fit of the data from the WFI camera on the 2.2-m MPG/ESO telescope. 

Finally, we included the BCG data of \citetalias{Sartoris2020}, which consist in a set of nine points of the VDP $[1,50]\,\text{kpc}$. The mass profile is modeled following the Jaffe profile, eq.~\eqref{eq:jaffe}, with a stellar mass of $M_{*} = 1.2^{+0.2}_{-0.6}\times 10^{12}\, \text{M}_\odot$ and a characteristic radius $r_j = 22.7 \pm 2.1 \text{kpc}$ (see Section \ref{sec:MAM}).

\citet{Mercurio_2021} performed a detailed analysis of the dynamical state of Abell~S1063, showing that the cluster underwent a recent off-axis merger, and it is not dynamically relaxed. This is further suggested by a non-Guassian velocity dispersion along the line of sight and by the comparison of the kinematic mass profile with the total mass profiles obtained from the Chandra X-ray data (see also \citealt{Bonamigo18}) and the strong+weak lensing analysis of \citet{Umetsu16} and \citet{Caminha16}. While the profiles are in overall good agreement, a discrepancy with the (non-parametric) weak-lensing results at $~0.3 $~Mpc has been found (see Figure~7 of \citetalias{Sartoris2020}). The possible un-relaxed state is further indicated by the analyses of \citet{Pizzuti17} and \cite{Pizzuti:2022Vain} in modified gravity scenarios.

\section{Mass modeling with \textsc{MG-MAMPOSSt}}\label{sec:MAM}
The \textsc{MG-MAMPOSSt} code (\citealt{Pizzuti:2022ynt}) is a version of the  \textsc{MAMPOSSt} (Modelling Anisotropy and Mass Profile of Spherical Observed Systems) method of \cite{Mamon01}.  Both codes jointly reconstruct the total gravitational potential and the orbit anisotropy profile of clusters using kinematic analyses of the member galaxies: \textsc{MAMPOSSt} in the framework of Newtonian gravity, whereas \textsc{MG-MAMPOSSt} extends the parametrisation of the gravitational potential to the framework of modified gravity models.

\textsc{MG-MAMPOSSt} performs a Maximum-Likelihood fit using data of the projected phase space $(R,v_z)$ of the cluster member galaxies  (p.p.s., hereafter), where $R$ is the projected position from the cluster center and $v_{z}$ is the l.o.s. velocity, computed in the rest frame of the cluster. The code assumes spherical symmetry, dynamical relaxation and  a Gaussian distribution of the 3D velocity field to solve the stationary spherical Jeans equation,
\begin{equation}
\label{eq:jeans}
\frac{\text{d} (\nu \sigma_r^2)}{\text{d} r}+2\beta(r)\frac{\nu\sigma^2_r}{r}=-\nu(r)\frac{\text{d} \Phi}{\text{d} r}\,,
\end{equation}
to derive 
the radial velocity dispersion $\sigma^2(r)$ as a function of the number density profile of the tracers $\nu(r)$, the velocity anisotropy profile $\beta \equiv 1-(\sigma_{\theta}^2+\sigma^2_{\varphi})/2\sigma^2_r$ and the gravitational potential $\Phi(r)$. In the definition of $\beta$, $\sigma_{\theta}^2,\,\sigma^2_{\varphi}$ represent the velocity dispersion along the tangential and azimuthal directions, respectively. Spherical symmetry enforces  $\sigma_{\theta}^2 = \sigma^2_{\varphi}$.

Given a parametric expression of $\Phi(r)\,, \beta(r)$ and $\nu(r)$, \textsc{MG-MAMPOSSt} estimate the (log) likelihood $\ln \mathcal{L}_{\text{dyn}}(\Theta)$ as the sum of the log probabilities of finding a galaxy at the p.p.s. position $(R_i, v_{z,i})$ for a radial velocity dispersion $\sigma^2(r,\Theta)$, described by the vector of parameters $\Theta$. 
The code allows for a quite broad range of parametrizations for the gravitational potential, including popular modified gravity / Dark Energy scenarios viable at the cosmological level, and very general models of the velocity anisotropy profile, which account for a wide variety of orbits of the member galaxies. Here we are going to apply \textsc{MG-MAMPOSSt} to the framework of RG, (eq.~\ref{eq:RG}); for a detailed description of the methods and the implemented physics, see \cite{Mamon01} and \cite{Pizzuti2021}.

We model the BCG stellar mass following the luminosity profile of eq.~(\ref{eq:jaffe}) as $M_{\text{BCG}}(r) = X_L\times L_J(r)$, where $X_L=M_*/L_{\text{BCG}}$ is the stellar mass-to-light ratio, which can be considered as a free parameter to be fitted in the \textsc{MG-MAMPOSSt} procedure. 
We fix $M_{\text{gas}}$ and $M_{\text{gal}}$ to the best fit profiles, as done in \citetalias{Biviano23}; while the uncertainties in the baryonic components provides negligible effects on the estimated profile in $\Lambda$CDM - where the dark matter component dominates, one may argue that in a model where the gravitational potential depends entirely on the baryons, this is no longer true. However, we checked that accounting for those uncertainties in the \textsc{MG-MAMPOSSt} fit does not produce any relevant change on the constrained RG parameters.

For the velocity anisotropy of galaxies we adopt a generalization of the Tiret model  \citep{Tiret2007}, gT hereafter  (see \citealt{Mamon19}):
\begin{equation}
    \beta = \beta_0 +(\beta_{\infty}-\beta_0) r/(r+r_{\beta})\,. \label{e:tiret}
\end{equation}
Here, $\beta_0$ and $\beta_\infty$ are the values of the anisotropy at the center and at large radii, respectively, and $r_{\beta}$ is a characteristic scale radius. In our analysis we use the scaled parameters $\mathcal{A}_{0/\infty} = (1 - \beta_{0/\infty})^{-1/2}$, which are equal to one for completely isotropic orbits. Moreover, for both clusters, we set $r_{\beta} = 0.5\,\text{Mpc}$, as we found that its variation over a broad range of values $[0,3]\,\text{Mpc}$ produces no significant improvement over the final posterior.
As for the BCG stellar velocity anisotropy profile, we consider for both clusters isotropic orbits, $\beta_{\text{BCG}}=0$, consistent with the results of \citetalias{Sartoris2020,Biviano23}. 

Note that in the analysis of \citetalias{Sartoris2020} the value of the stellar mass of the BCG is kept fixed to the best fit value obtained from fitting the SED, $M_{*} = 1.2^{+0.2}_{-0.6}\times 10^{12}\, \text{M}_\odot$. 
To facilitate comparisons and for consistency with the structure of our analysis, we consider $X_L(\text{Abell S1063)} = M_{*}/(3\times 10^{11}$ L$_\odot$), such that $X_L = 4.0$ M$_\odot$/L$_\odot$ for $M_{*} = 1.2 \times 10^{12}\, \text{M}_\odot $.
Moreover, \citetalias{Sartoris2020} assumed a Tiret profile for the velocity anisotropy of the member galaxies. Here we perform a preliminary run of \textsc{MG-MAMPOSSt} in Newtonian gravity 
for Abell S1063, following the same prescription as done in \citetalias{Biviano23} for the case of MACS 1206, and adopting a gT model for $\beta(r)$ (see Section \ref{sec:results}).

The number density distribution of the tracers $\nu(r)$ is determined directly from the projected phase space, taking into consideration the completeness of the sample. As done in previous works (e.g. \citetalias{Sartoris2020}; \citealt{Pizzuti:2022Vain}; \citetalias{Biviano23}) we perform a Maximum Likelihood fit to the numerical distribution of the member galaxies which does not require a binning of the data (\citealt{Sarazin80}) assuming a projected NFW model. Each galaxy is weighted by the inverse of the completeness within the fit. Since the normalization constant of $\nu(r)$ is not relevant in \textsc{MG-MAMPOSSt}, as it factors out in the solution of the Jeans equation, the only free parameter is the scale radius of the profile $r_\nu$. We obtain $r_\nu=0.46^{+0.08}_{-0.07}$ Mpc for MACS 1206 and $r_\nu=0.76^{+0.08}_{-0.05}$ Mpc for Abell S1063.
Following \citetalias{Biviano23}, in the \textsc{MG-MAMPOSSt} fit we let $r_\nu$ to be free, assuming flat priors within the $1\sigma$ uncertainties.

\section{Results}
\label{sec:results}
We run \textsc{MG-MAMPOSSt} considering the joint dataset of the member galaxies p.p.s. and the BCG stellar l.o.s. velocity dispersion profile. We compute the log likelihood as 
\begin{equation}
    \ln \mathcal{L}_{\text{tot}} =  
    \ln \mathcal{L}_{\text{pps}} - \frac{\chi^2_{\text{BCG}}}{2} \,, 
\end{equation}
where $\chi^2_{\text{BCG}}$ is the chi-square obtained by comparing the modeled 
$\sigma^{(P)}_{\text{BCG}}$ and the observed $\sigma^{(O)}_{\text{BCG}}$ l.o.s. velocity dispersion (see Equation 10 of \citetalias{Sartoris2020}).

For both clusters, our set of free parameters is given by $\Theta = \{ r_\nu,\mathcal{A}_0,\mathcal{A}_\infty, \epsilon_0, \log_{10}\,\rho_c, Q, X_L \}$. We perform a Monte-Carlo-Markov-Chain (MCMC) sampling of 100000 points in the parameter space assuming quite broad flat priors, listed in Table~\ref{tab:priors}.

\begin{table}
\begin{tabular}{ccc} \hline
parameter  & lower value	 & upper value	 \\
\midrule
$r_\nu$ [Mpc] (MACS 1206) &  0.39 & 0.54 \\
$r_\nu$ [Mpc] (Abell S1063) &  0.71 & 0.84 \\
$\mathcal{A}_0 $ & 0.5  & 4.5 \\
$\mathcal{A}_\infty $ & 0.5  & 4.5\\
$\epsilon_0$ & 0.05  & 0.30 \\ 
$\log_{10}\,(\rho_c\,[\text{M}_\odot\,\text{Mpc}^{-3}])$ &  13.0 & 16.0 \\
$Q$ & 0.1  & 30.0 \\
$X_L\,[\text{M}_\odot/\text{L}_\odot]$ & 0.2  & 7.0 \\
\bottomrule

\end{tabular}

\caption{\label{tab:priors}Prior ranges for the free parameters in the \textsc{MG-MAMPOSSt} analysis.}
\end{table}

\subsection{MACS 1206}

The results of the analysis of MACS1206 are presented in  Figure~\ref{f:macsContour} and in the first row of Table~\ref{tab:results}.\footnote{Note that we do not report the values of $r_\nu$ in Table~\ref{tab:results}, as they are not relevant for the results of our analysis.}

\begin{table*}
\centering
\begin{tabular}{cccccccc} \hline
Cluster  & $\mathcal{A}_0 $ & $\mathcal{A}_\infty $	 & $\epsilon_0 $ & $\log_{10}\,\rho_c\,[\text{M}_\odot\,\text{Mpc}^{-3}]$ & $Q$	 &  $X_L \,[\text{M}_\odot/\text{L}_\odot]$ & $\Delta$BIC	\\ \midrule
MACS 1206 (RG) & $1.38^{+1.42 }_{-0.95}$ &  $2.84^{+2.03 }_{-1.51}$ & $0.18^{+0.04 }_{-0.04}$ & $14.92^{+0.01 }_{-0.01}$ & $5.21^{+1.85 }_{-1.70}$ & $4.45^{+0.24 }_{-0.26}$ & $2.12$\\ \midrule
MACS 1206 (GR) & 
$ 1.65^{+0.83 }_{-0.65} $ & $3.19^{+1.72 }_{-1.59} $
& & & & $4.48^{+0.26 }_{-0.28}$ & 0\\
\midrule \midrule

Abell S1063 (RG) & $2.06^{+1.94 }_{-1.12}$ &  $1.01^{+0.31 }_{-0.27}$ & $0.13^{+0.01 }_{-0.01}$ & $14.787^{+0.002 }_{-0.003}$ & $23.62^{+5.00 }_{-4.69}$ & $4.88^{+0.28 }_{-0.27}$ & 4.1 \\ \midrule

Abell S1063 (GR) & 
$ 1.35^{+2.16 }_{-0.75} $ & $1.89^{+2.27 }_{-1.19} $
& & & & $4.15^{+0.44 }_{-0.47}$ & 0 \\

\bottomrule
\end{tabular}
\caption{\label{tab:results}Best fit parameters from the MCMC analysis of \textsc{MG-MAMPOSSt} for MACS 1206 (first row) and Abell S1063 (second row). The last column indicates the $\Delta$BIC with respect to the GR analysis of \citetalias{Biviano23} (MACS 1206) and to our preliminary run based on \citetalias{Sartoris2020} (Abell S1063), displayed in the third and fourth rows, respectively, for the parameters of interest. All the uncertainties are given at $95\%$ C.L.}
\end{table*}

In particular, Figure~\ref{f:macsContour} shows the marginalized posterior distributions of all the free parameters, and the two-dimensional contours at $1\,\sigma$ and $2\,\sigma$. In the last column of Table~\ref{tab:results} we have further computed the difference of the Bayesian Information Criterion (BIC) with respect to the 
 Newtonian result of \citetalias{Biviano23}.
The BIC (\citealt{Schwarz1978}) gives a fast estimation of which model performs better than the other; it is defined as $-2\ln \mathcal{L}_{\text{MLE}}+N_{\text{Par}}\ln\,N_{\text{Data}}$,
where $\mathcal{L}_{\text{MLE}}$ is the likelihood at the best fit, $N_{\text{Par}}$ the number of free parameters and $N_{\text{Data}}$ the number of data-points. A model is significantly better (worse) than another whenever $\Delta$BIC $\lesssim -6$ ($\gtrsim 6$).  

For MACS 1206, the Refracted Gravity framework performs adequately well as  Newtonian gravity, 
where the additional collisionless DM component is assumed. Moreover, the values of the common free parameters, namely  $\mathcal{A}_0,\, \mathcal{A}_\infty$, and the stellar mass-to-light ratio of the BCG $X_L$ 
are in agreement with the estimation of 
\citetalias{Biviano23}. 

\begin{figure*}
\centering
\includegraphics[width=0.9\textwidth]{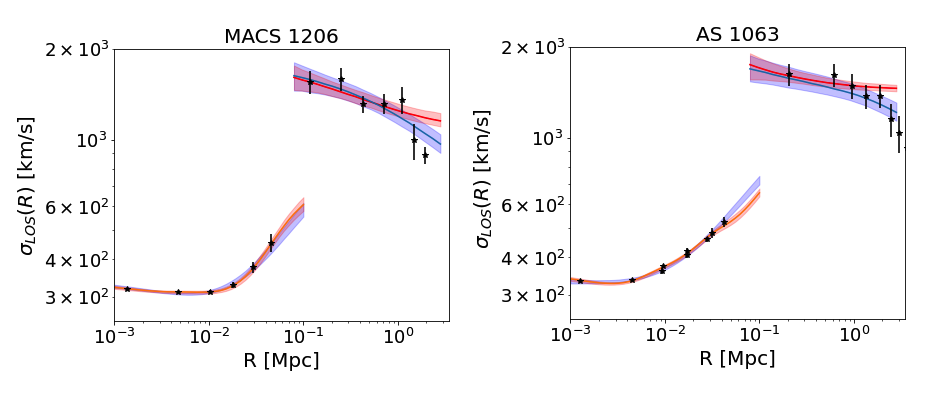}
\caption{Black points: observed line-of-sight velocity dispersions of the BCG and of the cluster, traced by member galaxies, for MACS 1206 (left) and Abell S1063 (right). The errorbars correspond to 1 $\sigma$ uncertainties. The red curves and shaded regions represent the best fit  profile in RG, and its 1$\sigma$ uncertainties, compared to the Newtonian results (blue shaded bands) of \citetalias{Biviano23} and \citetalias{Sartoris2020}, respectively.}
\label{fig:mass}
\end{figure*}

\begin{figure*}
\includegraphics[width=0.9\textwidth]{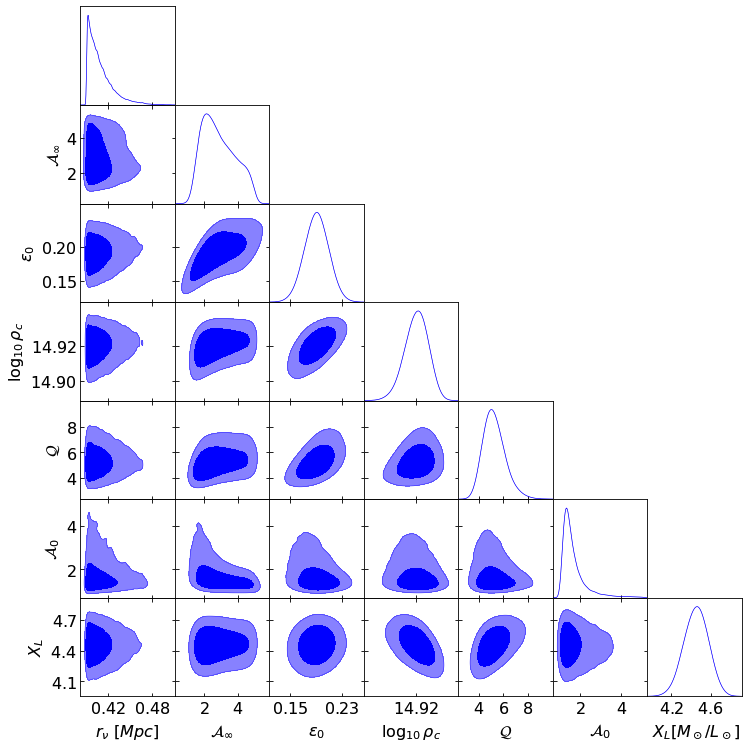}
\caption{Marginal two-dimensional and one-dimensional distribution for the free parameters of the \textsc{MG-MAMPOSSt} run in RG for MACS 1206. The darker and lighter coloured areas represents the 65\% and 95\% confidence regions, respectively. The critical density is in units of [M$_\odot$ Mpc$^{-3}$]. The plot has been obtained by means of the \href{https://getdist.readthedocs.io/en/latest/}{getdist} package.}
\label{f:macsContour}
\end{figure*}
\begin{figure}
\centering
\includegraphics[width=0.8\columnwidth]{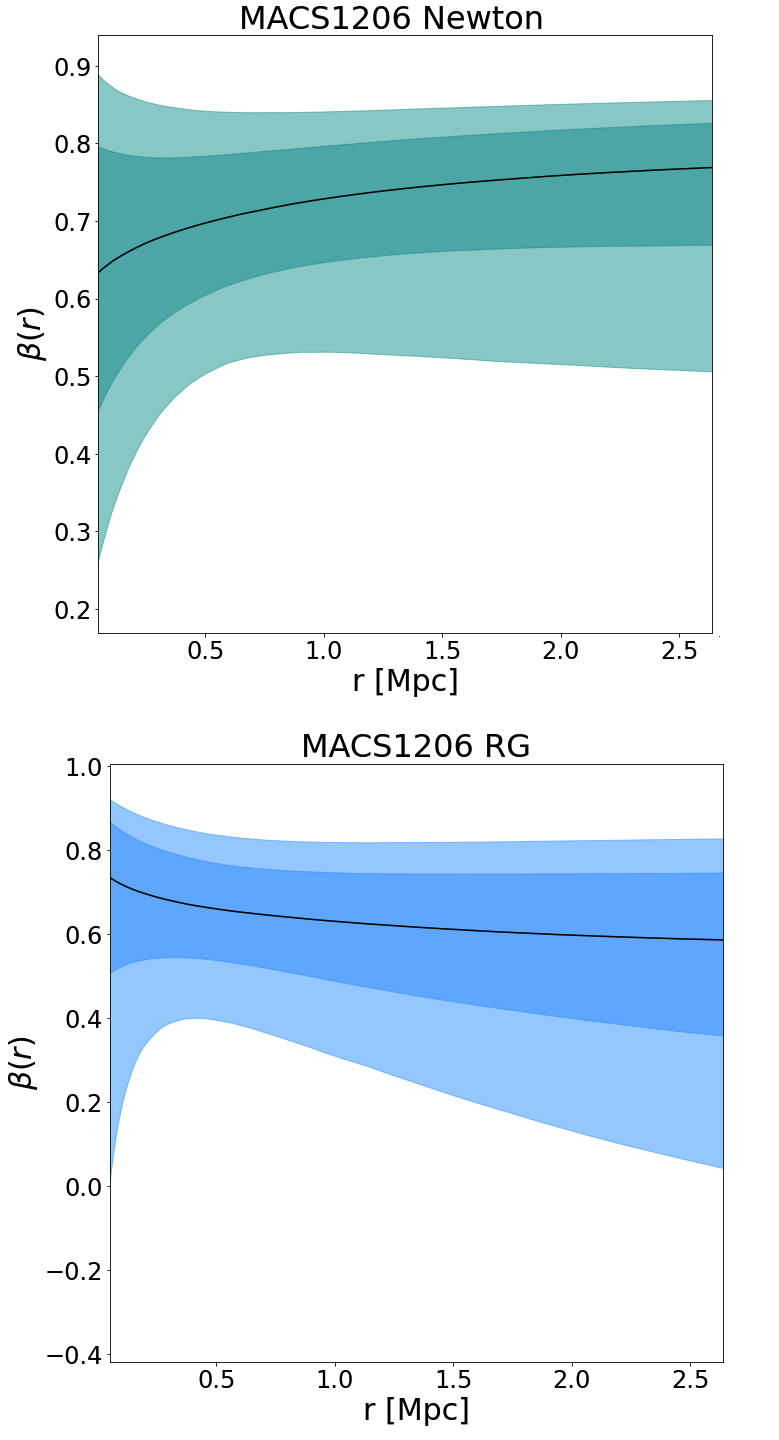}
\caption{Anisotropy profiles for MACS 1206 obtained by the \textsc{MG-MAMPOSSt} analysis in Newtonian gravity of \citetalias{Biviano23} (top) and RG (bottom). The inner and outer shaded regions represent the one and two $\sigma$ contours respectively.}
\label{fig:MACSanis}
\end{figure}

\subsection{Abell S1063}
In the case of Abell S1063, we first perform  a run in GR considering the same  priors as in Table \ref{tab:priors} for the common parameters $r_\nu,\,\mathcal{A}_0,\,\mathcal{A}_\infty,\,X_L$. As in \citetalias{Sartoris2020}, we adopt a generalized Navarro-Frenk-White (gNFW) model for the DM distribution characterized by three parameters, a "virial" radius $r_{200}$, a scale radius $r_\text{s}$ and an inner slope $\gamma$. We adopt the following flat priors for the parameters:
\begin{equation*}
   r_{200} \in [1.5,5.0]\,\text{Mpc},\,\,\,\, r_\text{s} \in [0.05,4.0]\,\text{Mpc},\,\,\, \gamma \in [0.01,2.0].
\end{equation*}
The results are in agreement with the findings of \citetalias{Sartoris2020}; for the DM mass profile parameters we obtain $r_{200} = 2.48^{+0.30 }_{-0.28}$ Mpc, $r_\text{s} = 0.71^{+0.32 }_{-0.27}$ Mpc, $\gamma = 0.95^{+0.15 }_{-0.17}$, where the uncertainites are given at $95\%$ C.L.
As for the anisotropy of member galaxies and the $X_L$ parameter of the BCG mass profile we find:
\begin{equation}\label{eq:betaGRAS}
\mathcal{A}_0 = 1.35^{+2.16 }_{-0.75} \,\,\,\,\,\,\, \mathcal{A}_\infty = 1.89^{+2.27 }_{-1.19} \,\,\,\,\, X_L = 4.15^{+0.44 }_{-0.47}\,\text{M}_\odot/\text{L}_\odot\,.
\end{equation}
Note that the constraint on $X_L$ can be translated to $M_* = 1.25^{+0.13}_{-0.14}\times 10^{12}\, M_\odot$, consistent with what found by the fit of the SED.

\begin{figure*}
\centering
\includegraphics[width=0.9\textwidth]{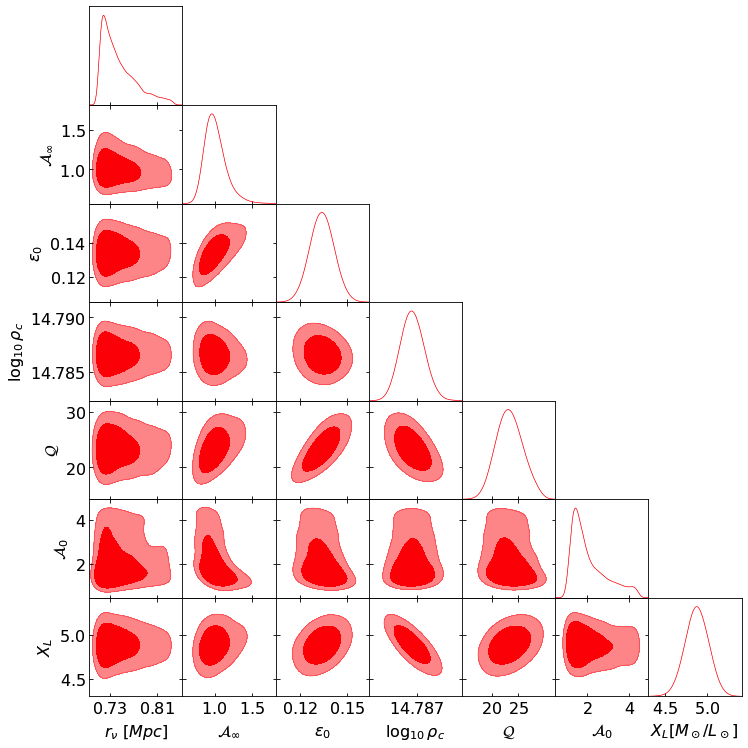}
\caption{Marginal two-dimensional and one-dimensional distribution for the free parameters of the \textsc{MG-MAMPOSSt} in RG for Abell S1063. The darker and lighter coloured areas represents the 65\% and 95\% confidence regions, respectively.  The critical density is in units of [M$_\odot$ Mpc$^{-3}$].}
\label{f:AbellContour}
\end{figure*}

\begin{figure}

\includegraphics[width=0.8\columnwidth]{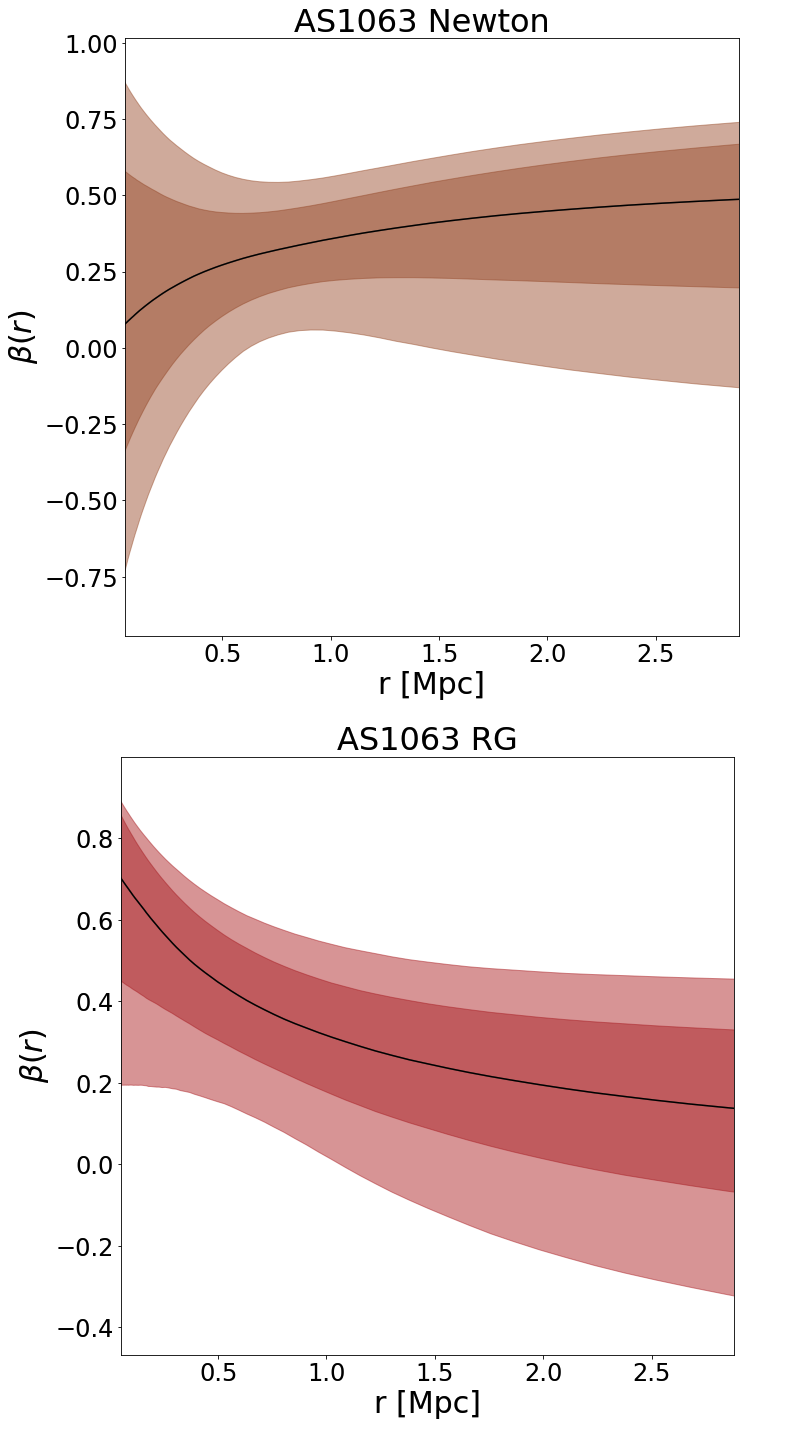}
\caption{Anisotropy profiles for AS 1063 obtained by our preliminary \textsc{MG-MAMPOSSt} analysis in Newtonian gravity (top) and RG (bottom). The inner and outer shaded regions represent the one and two $\sigma$ contours respectively.}
\label{fig:ASanis}
\end{figure}

 The marginalized distributions of the free parameters derived in the RG framework are shown in Figure \ref{f:AbellContour}. The values of the anisotropy parameters  $\mathcal{A}_0$ and $\mathcal{A}_\infty$ are in agreement with the  Newtonian 
result of eq. \eqref{eq:betaGRAS}, within the uncertainties. However, similarly to MACS 1206 (Figures \ref{fig:MACSanis}), $\beta(r)$ decreases with $r$ in RG and increases with $r$ in Newtonian gravity (Figure~\ref{fig:ASanis}): this different behavior is more evident in AS1603 than in MACS 1206.

As for the mass-to-light ratio parameter, the RG scenario seems to prefer larger values with respect to the result of the SED fitting (see the second row of Table \ref{tab:results}). The constraint on the stellar mass
$M_* = (1.46\pm 0.08)\times 10^{12}\, M_\odot$, is slightly in tension  ($\lesssim 2 \sigma$) with the quoted value in \citetalias{Sartoris2020}.

It is worth to notice that the preference of Newtonian gravity 
with respect to RG in this case is larger, $\Delta$BIC $= 4.2$, but still not statistically relevant. 

In the left panel of Figure \ref{fig:mass} we further show the  modeled 
velocity dispersion profile along the line-of-sight, $\sigma_{LOS}^{(P)}(R)$, of MACS 1206 (solid red line, light red shaded areas) in RG, compared with the observed one (black stars with errobars). The central region ($R\lesssim 0.1 \,\text{Mpc}$) refers to the stellar VDP of the BCG, while for $R>0.1 \,\text{Mpc}$ the profile of the member galaxies is presented. The blue curves represent $\sigma_{LOS}^{(P)}(R)$ for the Newtonian 
analysis of \citetalias{Biviano23}. In the right panel we plot the same for Abell S1063. 

For both clusters we find a very good agreement between the observed $\sigma_{LOS}(R)$ and the 
RG  model, except in the cluster outskirts, where the RG profiles start to flatten, deviating from the observed trend. This result can be caused by several factors; first, the estimate of the velocity dispersion profile  in the external regions of galaxy clusters may suffer for a larger contamination from interlopers, which can produce an underestimate 
of the observed value of $\sigma_{LOS}(R)$ (see e.g. \citealt{mamon10}). Moreover, at large radii, increasing lack of spherical symmetry and dynamical relaxation may affect the kinematic analysis - which is based on the Jeans equation - producing a bias in the velocity dispersion profile reconstruction. RG is particularly sensitive to the failure of these assumptions, because, 
as mentioned in Section \ref{sec:theo}, the gravitational field lines are significantly modified by the shape of the system. We will further discuss this point in the next Section.

\section{Discussion}
    \label{sec:disc}
    In the following we compare the results obtained from our analysis of the two clusters with existing constraints on the RG parameters in the literature. In particular, we consider  the results from the study of disk and E0 elliptical galaxies by \cite{RG2020, RG2023}, listed in the first and second columns of Table~\ref{tab:RG_review}. To provide a quantitative estimate of the discrepancies among the parameters, we compute the so-called sigma distance, defined as
    $ d = {|A - B|}/{\sigma}$. 
    $A$ and $B$ are two best fit values of the same parameter for different estimations (the analyses of the two clusters), and $\sigma=\sqrt{\sigma_A^2+\sigma_B^2}$, with  $\sigma_A$ and $\sigma_B$  the uncertainties on the individual parameters. 
    For a-symmetric distributions,  we compute $\sigma_{A,B}$ in the following way:
    if $A > B$ we consider $\sigma_A = \sigma_A^{(-)}$ and $\sigma_B = \sigma_B^{(+)}$, where $\sigma^{(-)}$, $\sigma^{(+)}$ are the differences between the best fit and the 16$^{(th)}$ percentile, and the difference between the 84$^{(th)}$ percentile and the best fit, respectively. The opposite for $A < B$.
    The purpose is to compare the closest sides of the A and B distributions

\begin{table*}
\centering
\begin{tabular}{ccccc} \hline
parameter  & Disk galaxies	 & Elliptical galaxies	&  MACS 1206 & Abell S1063\\
\midrule
\vspace{3pt}
$\log_{10}\left[\rho_c/(\text{g}_\odot/\text{cm}^3)\right]$  & $-24.54^{+0.08}_{-0.07}$  & $-24.25^{+0.28}_{-0.20}$ & $-25.25^{+0.01}_{-0.01}$ & $-25.383^{+0.003}_{-0.002}$\\ \vspace{3pt}
$\epsilon_0$  &  $0.666^{+0.007}_{-0.007}$ & $0.089^{+0.038}_{-0.035}$ & $0.18^{+0.04 }_{-0.04}$ & $0.13^{+0.01 }_{-0.01}$\\
$Q $ & $1.79^{+0.14}_{-0.26}$  & $0.47^{+0.29}_{-0.21}$ & $5.21^{+1.85 }_{-1.70}$ & $23.62^{+5.00 }_{-4.69}$\\
\bottomrule
\end{tabular}
\caption{\label{tab:RG_review}Comparison of the constraints on the RG parameters of this work (third and fourth column) and the results from the analyses of \cite{RG2020,RG2023} (first and second column).}
\end{table*}
\textit{MACS1206:} 
the estimated critical density parameter $\log_{10} \rho_c$  differs by about $5 \,\sigma$ and $10 \,\sigma$ from the value of the E0 ellipticals and the disk galaxies, respectively. The other two parameters are in poor agreement: the value of $\epsilon_0$ obtained for MACS1206 is $\sim 2\, \sigma$ away from the elliptical value and $12 \,\sigma$ from the result of the disk galaxies; for the slope parameter $Q$ we obtained a value 2.7 $\sigma$ away from the elliptical one and 2 $\sigma$ from the result of disk galaxies. \\

\textit{Abell S1063:} The constraints on the RG parameters differ significantly from the one obtained at galaxy scale, both for the elliptical and disk galaxies: the critical density parameter $\log_{10}\rho_c$ is $10\, \sigma$ away form the value of the disk galaxies and $6 \sigma$ from the result of the ellipticals;  $\epsilon_0$ is more than 40 $\sigma$ smaller than the value found for disk galaxies, whereas 
it is $1 \, \sigma$ away from the elliptical value. The slope parameter $Q$ 
is $4.6 \, \sigma$ away from the disk galaxy result and $5 \, \sigma$ from the elliptical value. \\

\begin{figure*}
\includegraphics[width=0.9\textwidth]{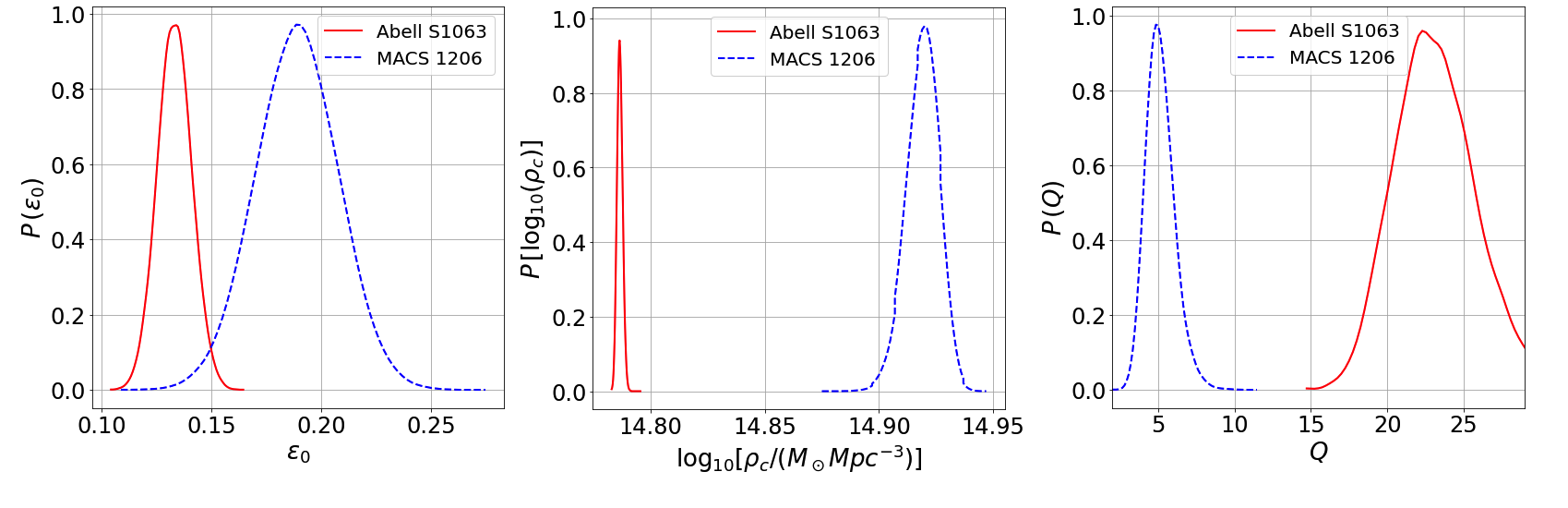}
\caption{Marginal one-dimensional distributions of the RG  parameters obtained from the \textsc{MG-MAMPOSSt} analyses for MACS 1206 (blue dotted curves) and Abell S1063 (red solid curve).}
\label{fig:1dim}
\end{figure*}

In Figure \ref{fig:1dim}, we further show the one-dimensional marginalized distributions of $\epsilon_0,\,\log_{10}\rho_c$ and $Q$ for MACS 1206 (blue dotted line) and Abell S1063 (red solid line). While the discrepancy between the values of $\epsilon_0$ is not relevant ($\sim 1 \sigma$), the tension is pretty large in the case of the other two parameters: $13 \sigma$ for the critical density,  and $4 \sigma$ for the slope $Q$.

The tension between the two sets of RG parameters is confirmed by the following  additional check: we performed a run of Abell S1063 in RG but using as prior for the RG parameters, $\epsilon_0$, $\log_{10}\,\rho_c$ and $Q$, the values obtained in the analysis of MACS 1206 (first row of Table \ref{tab:results}). In this case, even though the $\Delta$BIC with respect to the GR analysis is slightly lower $\Delta$BIC$=3.23$, the constraints on the stellar mass-to-light ratio, $X_L=1.15^{+0.15}_{-0.13}$ M$_\odot$/L$_\odot$, are completely inconsistent with the SED fitting of \citetalias{Sartoris2020}.

It is interesting to note that the tension between the constraints is reduced when comparing spherically-symmetric structures (i.e. elliptical and clusters); this may be a consequence of the rather different phenomenology in RG for flat systems and spherically symmetric halos (e.g \citealt{matsakos2016dynamics,Cesare2024}).

As pointed out in \cite{Cesare2024}, the  discrepancies between the observed values of the RG parameters can be due to several causes. At first, fixing the gas and galaxies mass components for both clusters can result in overfitting the posterior from \textsc{MG-MAMPOSSt}; nevertheless, as mentioned in Section \ref{sec:MAM}, even when accounting for the uncertainties in the baryonic mass profiles the tension is not reduced. Secondly, the inconsistencies on the values of the vacuum permittivity $\epsilon_0$ can  originate 
by the fact that 
clusters and elliptical galaxies 
have been approximated as isolated systems.  However, elliptical galaxies are generally located in group and clusters, and galaxy clusters are the nodes of the large  scale distribution. Environmental effects may have a relevant effect on the RG gravitational field  
which, as discussed, depends on the local density. 

More importantly, possible departures from spherical symmetry could introduce biases in the obtained constraints; as shown in e.g. \cite{Pizzuti19b},  observing a prolate halo along the major axis causes an overestimate of the mass, reflecting in spurious detection of modified gravity even for clusters in $\Lambda$CDM. This is particularly relevant in RG, where asphericity plays a crucial role in enhancing the modification of gravity (see e.g. \citealt{Cesare2024}): non-spherical, flattened systems, such as disk galaxies, exhibit more pronounced gravitational refraction effects compared to spherical systems due to the  redirection of the field lines (\citealt{matsakos2016dynamics}).

 The analysis of a much larger number of systems at different scales is required 
to provide a more solid conclusion 
about the viability of the RG model. 


\section{Conclusions}\label{sec:conc}
In this paper we have reconstructed the total mass profile of two galaxy clusters, MACS 1206 and Abell S1063, from kinematic analyses in the framework of RG (\citealt{matsakos2016dynamics}). In particular, we have analysed high precision data of the projected phase space of the member galaxies and of the stellar velocity dispersion of the BCG by means of the \textsc{MG-MAMPOSSt} method. Through an accurate modeling of the baryonic components in each cluster, we constrained the free parameters of the phenomenological RG theory, namely the  vacuum
gravitational permittivity $\epsilon_0$, the critical density $\rho_c$ and the slope $Q$, which are supposed to be universal.

Our analysis indicates that the RG framework can efficiently recover the total gravitational potential of a cluster-size structure assuming that the only mass component is given by baryons; the model provides an adequate fit to the kinematic data which is not strongly disfavoured with respect to the $\Lambda$CDM scenario. In particular we found $\Delta$BIC $= 1.1, \,\, 4.2$ for MACS 1206 and Abell S1063, respectively.
However, the constraints on the model's free parameters appear to be very different between the two clusters; while the tension is mild for the  vacuum gravitational permittivity ($< 2\sigma$) it becomes very strong ($> 12\sigma$) for the critical density. The tension is further exacerbated 
by the comparison with the other existing constraints on the  RG parameters obtained by the analysis of disk and E0 elliptical galaxies (\citealt{RG2020, RG2023}).   Although this tension might naively  suggest 
a lack of universality of the parameters, 
the serious discrepancies may be a consequence of the several assumptions made in the analyses. In particular, we have considered both clusters as spherically symmetric, isolated and relaxed systems; the lack of spherical symmetry, which may become larger in the cluster outskirt (e.g. \citealt{Morandi2015TheGC}), produces a strong impact in the RG framework, biasing the obtained constraints when not accurately taken into account. Similarly, environmental effects and departures from dynamical equilibrium are likely to have relevant effects on the gravitational field which is generated by the baryonic matter alone. 
A different origin of the discrepancies might be that the adopted functional form of the permittivity $\epsilon(\rho)$ is inappropriate and/or that $\epsilon$ depends on other local properties in addition to the density of the sources of the gravitational field.


It is worth to point out that the kinematic reconstruction performed here is based on the Jeans equation, which further requires that the cluster is a dynamically relaxed system. While relaxation is a valid assumption for MACS 1206, this is not the case of Abell S1063, as discussed in Section \ref{sec:data}. The more disturbed state of the latter may partially reflect on the tension highlighted by our analysis. Nevertheless, it is interesting to mention that a recent study (\citealt{ferrami2023dynamics}) found the presence of mass segregation in both clusters without any evidence of energy equipartition, which suggests that this mass segregation is unrelated to collision events between member galaxies. This seems to indicate that galaxies can still be used as collisionless tracers of the gravitational potential, supporting the validity of the Jeans' approach.

A more intriguing possibility is that the phenomenological approach employed here could be  inappropriate to describe the kinematic of systems at all scales and environments. 
The covariant formulation of RG \citep{RGcovariant}, where a scalar field coincides with the gravitational permittivity in the weak field limit, is a non-linear theory that might produce gravitational effects that are unexpected in a naively derived weak field limit, similarly to what occurs in General Relativity and Newtonian gravity (see e.g., \citealt{Balasin08}). Therefore the tensions we find here might suggest that a more careful weak field limit of the covariant RG theory should be derived to investigate the dynamics of galaxies and galaxy clusters. 

In conclusion, RG offers an interesting alternative scenario to the $\Lambda$CDM paradigm, with a rich phenomenology able to reproduce the observed velocity dispersion profiles in galaxy clusters. However, the manifested inconsistency of the parameters found in our analysis points out to the necessity toWe also acknowledge partial support from the INFN
grant InDark. This research carefully model possible systematic effects, such as departure from dynamical relaxation and asphericity. Therefore, more tests at clusters scales are required, along with a deeper exploration and refinement of the theoretical features of the model. 


\begin{acknowledgements}
\label{sec:acknowledgements}
This study is supported by the Italian Ministry for Research and University (MUR) under Grant 'Progetto Dipartimenti di Eccellenza 2023-2027' (BiCoQ). AB acknowledges the financial contribution from the INAF mini-grant 1.05.12.04.01 {\it "The dynamics of clusters of galaxies from the projected phase-space distribution of cluster galaxies".} AD acknowledges partial support from the INFN
grant InDark. This research has made use of NASA’s Astrophysics Data System
Bibliographic Services. The authors acknowledge P. Rosati and the CLASH-VLT team for the dataset used in this work.  
\end{acknowledgements}

\citestyle{aa}
\bibliographystyle{aa}
\bibliography{sample631}

\begin{thebibliography}{54}
\expandafter\ifx\csname natexlab\endcsname\relax\def\natexlab#1{#1}\fi

\bibitem[{{Abell} {et~al.}(1989){Abell}, {Corwin}, \& {Olowin}}]{Abell89}
{Abell}, G.~O., {Corwin}, Jr., H.~G., \& {Olowin}, R.~P. 1989, \apjs, 70, 1

\bibitem[{Annunziatella {et~al.}(2014)Annunziatella, Biviano, Mercurio, Nonino,
  Rosati, Balestra, Presotto, Girardi, Gobat, Grillo, Kelson, Medezinski,
  Postman, Scodeggio, Brescia, Demarco, Fritz, Koekemoer, Lemze, Lombardi,
  Sartoris, Umetsu, Vanzella, Bradley, Coe, Donahue, Infante, Kuchner, Maier,
  Regős, Verdugo, \& Ziegler}]{Annunziatella2014}
Annunziatella, M., Biviano, A., Mercurio, A., {et~al.} 2014, A\&A, 571, A80

\bibitem[{Arbey \& Mahmoudi(2021)}]{Arbey2021}
Arbey, A. \& Mahmoudi, F. 2021, Progress in Particle and Nuclear Physics, 119,
  103865

\bibitem[{{Balasin} \& {Grumiller}(2008)}]{Balasin08}
{Balasin}, H. \& {Grumiller}, D. 2008, International Journal of Modern Physics
  D, 17, 475

\bibitem[{Balestra {et~al.}(2016)Balestra, Mercurio, Sartoris, Girardi, Grillo,
  Nonino, Rosati, Biviano, Ettori, Forman, Jones, Koekemoer, Medezinski,
  Merten, Ogrean, Tozzi, Umetsu, Vanzella, van Weeren, Zitrin, Annunziatella,
  Caminha, Broadhurst, Coe, Donahue, Fritz, Frye, Kelson, Lombardi, Maier,
  Meneghetti, Monna, Postman, Scodeggio, Seitz, \& Ziegler}]{Balestra_2016}
Balestra, I., Mercurio, A., Sartoris, B., {et~al.} 2016, The Astrophysical
  Journal Supplement Series, 224, 33

\bibitem[{Banik {et~al.}(2023)Banik, Pittordis, Sutherland, Famaey, Ibata,
  Mieske, \& Zhao}]{Banik_2023}
Banik, I., Pittordis, C., Sutherland, W., {et~al.} 2023, MNRAS, 527,
  4573–4615

\bibitem[{Biviano {et~al.}(2023)Biviano, Pizzuti, Mercurio, Sartoris, Rosati,
  Ettori, Girardi, Grillo, Caminha, \& Nonino}]{Biviano23}
Biviano, A., Pizzuti, L., Mercurio, A., {et~al.} 2023, ApJ, 958, 148

\bibitem[{{Biviano} {et~al.}(2021){Biviano}, {van der Burg}, {Balogh},
  {Munari}, {Cooper}, {De Lucia}, {Demarco}, {Jablonka}, {Muzzin}, {Nantais},
  {Old}, {Rudnick}, {Vulcani}, {Wilson}, {Yee}, {Zaritsky}, {Cerulo}, {Chan},
  {Finoguenov}, {Gilbank}, {Lidman}, {Pintos-Castro}, \& {Shipley}}]{Biviano21}
{Biviano}, A., {van der Burg}, R.~F.~J., {Balogh}, M.~L., {et~al.} 2021, A\&A,
  650, A105

\bibitem[{Bonamigo {et~al.}(2018)Bonamigo, Grillo, Ettori, Caminha, Rosati,
  Mercurio, Munari, Annunziatella, Balestra, \& Lombardi}]{Bonamigo18}
Bonamigo, M., Grillo, C., Ettori, S., {et~al.} 2018, ApJ, 864, 98

\bibitem[{Boumechta {et~al.}(2023)Boumechta, Haridasu, Pizzuti, Butt,
  Baccigalupi, \& Lapi}]{Boumechta:2023qhd}
Boumechta, Y., Haridasu, B.~S., Pizzuti, L., {et~al.} 2023, Phys. Rev. D, 108,
  044007

\bibitem[{Braglia {et~al.}(2021)Braglia, Ballardini, Finelli, \&
  Koyama}]{Braglia_2021}
Braglia, M., Ballardini, M., Finelli, F., \& Koyama, K. 2021, Phys. Rev. D, 103

\bibitem[{{Caminha} {et~al.}(2016){Caminha}, {Grillo}, {Rosati}, {Balestra},
  {Karman}, {Lombardi}, {Mercurio}, {Nonino}, {Tozzi}, {Zitrin}, \&
  {others}}]{Caminha16}
{Caminha}, G.~B., {Grillo}, C., {Rosati}, P., {et~al.} 2016, \aap, 587, A80

\bibitem[{{Caminha} {et~al.}(2017){Caminha}, {Grillo}, {Rosati}, {Meneghetti},
  {Mercurio}, {Ettori}, {Balestra}, {Biviano}, {Umetsu}, {Vanzella},
  {Annunziatella}, {Bonamigo}, {Delgado-Correal}, {Girardi}, {Lombardi},
  {Nonino}, {Sartoris}, {Tozzi}, {Bartelmann}, {Bradley}, {Caputi}, {Coe},
  {Ford}, {Fritz}, {Gobat}, {Postman}, {Seitz}, \& {Zitrin}}]{Caminha2017}
{Caminha}, G.~B., {Grillo}, C., {Rosati}, P., {et~al.} 2017, \aap, 607, A93

\bibitem[{{Cesare}(2024)}]{Cesare2024}
{Cesare}, V. 2024, Astronomy, 3, 68

\bibitem[{{Cesare} {et~al.}(2022){Cesare}, {Diaferio}, \& {Matsakos,
  T.}}]{RG2023}
{Cesare}, V., {Diaferio}, A., \& {Matsakos, T.} 2022, A\&A, 657, A133

\bibitem[{{Cesare} {et~al.}(2020){Cesare}, {Diaferio}, {Matsakos, T.}, \&
  {Angus, G.}}]{RG2020}
{Cesare}, V., {Diaferio}, A., {Matsakos, T.}, \& {Angus, G.} 2020, A\&A, 637,
  A70

\bibitem[{Dodelson {et~al.}(1996)Dodelson, Gates, \& Turner}]{Dodelson96}
Dodelson, S., Gates, E.~I., \& Turner, M.~S. 1996, Science, 274, 69–75

\bibitem[{{Ettori} {et~al.}(2010){Ettori}, {Gastaldello}, {Leccardi},
  {Molendi}, {Rossetti}, {Buote}, \& {Meneghetti}}]{Ettori10}
{Ettori}, S., {Gastaldello}, F., {Leccardi}, A., {et~al.} 2010, \aap, 524, A68

\bibitem[{Famaey \& McGaugh(2012)}]{Famaey_2012}
Famaey, B. \& McGaugh, S.~S. 2012, Living Reviews in Relativity, 15

\bibitem[{Ferrami {et~al.}(2023)Ferrami, Bertin, Grillo, Mercurio, \&
  Rosati}]{ferrami2023dynamics}
Ferrami, G., Bertin, G., Grillo, C., Mercurio, A., \& Rosati, P. 2023, A\&A,
  676, A66

\bibitem[{{Girardi} {et~al.}(2015){Girardi}, {Mercurio}, {Balestra}, {Nonino},
  {Biviano}, {Grillo}, {Rosati}, {Annunziatella}, {Demarco}, {Fritz}, \&
  {others}}]{girardi15}
{Girardi}, M., {Mercurio}, A., {Balestra}, I., {et~al.} 2015, \aap, 579, A4

\bibitem[{{Hodson} \& {Zhao}(2017)}]{Hodson17}
{Hodson}, A.~O. \& {Zhao}, H. 2017, \aap, 598, A127

\bibitem[{Joyce {et~al.}(2016)Joyce, Lombriser, \& Schmidt}]{Joyce16}
Joyce, A., Lombriser, L., \& Schmidt, F. 2016, Annual Review of Nuclear and
  Particle Science, 66, 95

\bibitem[{{Lemze} {et~al.}(2013){Lemze}, {Postman}, {Genel}, {Ford},
  {Balestra}, {Donahue}, {Kelson}, {Nonino}, {Mercurio}, {Biviano}, \&
  {others}}]{Lemze01}
{Lemze}, D., {Postman}, M., {Genel}, S., {et~al.} 2013, \apj, 776, 91

\bibitem[{{Limousin} {et~al.}(2022){Limousin}, {Beauchesne}, \&
  {Jullo}}]{Limousin22}
{Limousin}, M., {Beauchesne}, B., \& {Jullo}, E. 2022, \aap, 664, A90

\bibitem[{{Mamon} {et~al.}(2013){Mamon}, {Biviano}, \& {Bou{\'e}}}]{Mamon01}
{Mamon}, G.~A., {Biviano}, A., \& {Bou{\'e}}, G. 2013, \mnras, 429, 3079

\bibitem[{{Mamon} {et~al.}(2010){Mamon}, {Biviano}, \& {Murante}}]{mamon10}
{Mamon}, G.~A., {Biviano}, A., \& {Murante}, G. 2010, \aap, 520, A30

\bibitem[{{Mamon} {et~al.}(2019){Mamon}, {Cava}, {Biviano}, {Moretti},
  {Poggianti}, \& {Bettoni}}]{Mamon19}
{Mamon}, G.~A., {Cava}, A., {Biviano}, A., {et~al.} 2019, \aap, 631, A131

\bibitem[{{Matsakos} \& {Diaferio}(2016)}]{matsakos2016dynamics}
{Matsakos}, T. \& {Diaferio}, A. 2016, arXiv e-prints, arXiv:1603.04943

\bibitem[{Mercurio {et~al.}(2021)Mercurio, Rosati, Biviano, Annunziatella,
  Girardi, Sartoris, Nonino, Brescia, Riccio, Grillo, Balestra, Caminha,
  De~Lucia, Gobat, Seitz, Tozzi, Scodeggio, Vanzella, Angora, Bergamini,
  Borgani, Demarco, Meneghetti, Strazzullo, Tortorelli, Umetsu, Fritz, Gruen,
  Kelson, Lombardi, Maier, Postman, Rodighiero, \& Ziegler}]{Mercurio_2021}
Mercurio, A., Rosati, P., Biviano, A., {et~al.} 2021, A\&A, 656, A147

\bibitem[{{Milgrom}(1983)}]{Milgrom1983}
{Milgrom}, M. 1983, \apj, 270, 365

\bibitem[{Morandi {et~al.}(2015)Morandi, Sun, Forman, \&
  Jones}]{Morandi2015TheGC}
Morandi, A., Sun, M., Forman, W.~R., \& Jones, C. 2015, MNRAS, 450, 2261

\bibitem[{Perivolaropoulos \& Skara(2022)}]{Perivolaropoulos_2022}
Perivolaropoulos, L. \& Skara, F. 2022, New Astronomy Reviews, 95, 101659

\bibitem[{{Perlmutter} {et~al.}(1999){Perlmutter}, {Aldering}, {Goldhaber},
  {Knop}, {Nugent}, {Castro}, {Deustua}, {Fabbro}, \& {The Supernova Cosmology
  Project}}]{perlmutter99}
{Perlmutter}, S., {Aldering}, G., {Goldhaber}, G., {et~al.} 1999, \apj, 517,
  565

\bibitem[{{Pizzuti} {et~al.}(2021){Pizzuti}, {Saltas}, \&
  {Amendola}}]{Pizzuti2021}
{Pizzuti}, L., {Saltas}, I.~D., \& {Amendola}, L. 2021, MNRAS, 506, 595

\bibitem[{Pizzuti {et~al.}(2023)Pizzuti, Saltas, Biviano, Mamon, \&
  Amendola}]{Pizzuti:2022ynt}
Pizzuti, L., Saltas, I.~D., Biviano, A., Mamon, G., \& Amendola, L. 2023, JOSS,
  8, 4800

\bibitem[{Pizzuti {et~al.}(2022)Pizzuti, Saltas, Umetsu, \&
  Sartoris}]{Pizzuti:2022Vain}
Pizzuti, L., Saltas, I.~D., Umetsu, K., \& Sartoris, B. 2022, MNRAS, 512, 4280

\bibitem[{{Pizzuti} {et~al.}(2017){Pizzuti}, {Sartoris}, {Amendola}, {Borgani},
  {Biviano}, {Umetsu}, {Mercurio}, {Rosati}, {Balestra}, {Caminha}, {Girardi},
  {Grillo}, \& {Nonino}}]{Pizzuti17}
{Pizzuti}, L., {Sartoris}, B., {Amendola}, L., {et~al.} 2017, JCAP, 7, 023

\bibitem[{Pizzuti {et~al.}(2020)Pizzuti, Sartoris, Borgani, \&
  Biviano}]{Pizzuti19b}
Pizzuti, L., Sartoris, B., Borgani, S., \& Biviano, A. 2020, JCAP, 04, 024

\bibitem[{{Planck Collaboration}(2020)}]{Planck2020}
{Planck Collaboration}. 2020, \aap, 641, A1

\bibitem[{{Postman} {et~al.}(2012)}]{Postman12}
{Postman}, M. {et~al.} 2012, \apjs, 199, 25

\bibitem[{{Riess} {et~al.}(1998){Riess}, {Filippenko}, {Challis},
  {Clocchiatti}, {Diercks}, {Garnavich}, {Gilliland}, {Hogan}, {Jha},
  {Kirshner}, \& {others}}]{Reiss01}
{Riess}, A.~G., {Filippenko}, A.~V., {Challis}, P., {et~al.} 1998, \aj, 116,
  1009

\bibitem[{{Rosati} {et~al.}(2014){Rosati}, {Balestra}, {Grillo}, {Mercurio},
  {Nonino}, {Biviano}, {Girardi}, {Vanzella}, \& {Clash-VLT Team}}]{Rosati14}
{Rosati}, P., {Balestra}, I., {Grillo}, C., {et~al.} 2014, The Messenger, 158,
  48

\bibitem[{Salucci(2019)}]{Salucci2019}
Salucci, P. 2019, The Astronomy and Astrophysics Review, 27

\bibitem[{Sanna {et~al.}(2023)Sanna, Matsakos, \& Diaferio}]{RGcovariant}
Sanna, A.~P., Matsakos, T., \& Diaferio, A. 2023, A\&A, 674, A209

\bibitem[{{Sarazin}(1980)}]{Sarazin80}
{Sarazin}, C.~L. 1980, \apj, 236, 75

\bibitem[{{Sartoris} {et~al.}(2014){Sartoris}, {Biviano}, {Rosati}, {Borgani},
  {Umetsu}, {Bartelmann}, {Girardi}, {Grillo}, {Lemze}, {Zitrin}, \&
  {others}}]{SartorisDM}
{Sartoris}, B., {Biviano}, A., {Rosati}, P., {et~al.} 2014, \apjl, 783, L11

\bibitem[{{Sartoris} {et~al.}(2020){Sartoris}, {Biviano}, {Rosati}, {Mercurio},
  {Grillo}, {Ettori}, {Nonino}, {Umetsu}, {Bergamini}, {Caminha}, \&
  {Girardi}}]{Sartoris2020}
{Sartoris}, B., {Biviano}, A., {Rosati}, P., {et~al.} 2020, \aap, 637, A34

\bibitem[{Schwarz(1978)}]{Schwarz1978}
Schwarz, G. 1978, The Annals of Statistics, 6, 461

\bibitem[{Shankaranarayanan \& Johnson(2022)}]{Shankaranarayanan_2022}
Shankaranarayanan, S. \& Johnson, J.~P. 2022, General Relativity and
  Gravitation, 54

\bibitem[{Tereno {et~al.}(2011)Tereno, Semboloni, \& Schrabback}]{Tereno_2011}
Tereno, I., Semboloni, E., \& Schrabback, T. 2011, A\&A, 530, A68

\bibitem[{{Tiret} {et~al.}(2007){Tiret}, {Combes}, {Angus}, {Famaey}, \&
  {Zhao}}]{Tiret2007}
{Tiret}, O., {Combes}, F., {Angus}, G.~W., {Famaey}, B., \& {Zhao}, H.~S. 2007,
  \aap, 476, L1

\bibitem[{{Umetsu} {et~al.}(2012){Umetsu}, {Medezinski}, {Nonino}, {Merten},
  {Zitrin}, {Molino}, {Grillo}, {Carrasco}, {Donahue}, {Mahdavi}, \&
  {others}}]{UmetsuMACS}
{Umetsu}, K., {Medezinski}, E., {Nonino}, M., {et~al.} 2012, \apj, 755, 56

\bibitem[{{Umetsu} {et~al.}(2016){Umetsu}, {Zitrin}, {Gruen}, {Merten},
  {Donahue}, \& {Postman}}]{Umetsu16}
{Umetsu}, K., {Zitrin}, A., {Gruen}, D., {et~al.} 2016, \apj, 821, 116

\end{thebibliography}



\end{document}